\documentclass[a4paper,onecolumn,preprintnumbers,notitlepage,nofootinbib]{revtex4-1}
\usepackage[sumlimits,intlimits,namelimits]{amsmath}
\usepackage{amssymb}
\usepackage[english]{babel}
\usepackage{braket}
\usepackage{graphicx}
\usepackage{slashed}
\usepackage{xcolor}

\usepackage{soul}

\newcommand{\mLamb}{M_{\Lambda_b}}
\newcommand{\mcMSbar}{\overline{m}_c(\overline{m}_c)}
\newcommand{\mbMSbar}{\overline{m}_b(\overline{m}_b)}
\newcommand{\eps}{\varepsilon}

\newcommand{\GeV}{\text{GeV}}

\newcommand{\refeq}[1]{eq.~(\ref{eq:#1})}
\newcommand{\reffig}[1]{figure~\ref{fig:#1}}
\newcommand{\reftab}[1]{table~\ref{tab:#1}}
\newcommand{\refsec}[1]{section~\ref{sec:#1}}
\newcommand{\order}[1]{\mathcal{O}(#1)}

\begin{document}

\title{Zero-Recoil Sum Rules for \boldmath $\Lambda_b \to \Lambda_c$ \unboldmath Form Factors}
\author{Thomas Mannel}
\email{mannel@physik.uni-siegen.de}
\author{Danny van Dyk}
\email{vandyk@physik.uni-siegen.de}
\affiliation{Theoretische Physik 1, Naturwissenschaftlich-Technische Fakult\"at, Universit\"at Siegen,
Walter-Flex-Straße 3, D-57068 Siegen, Germany}
\preprint{SI-HEP-2015-14, QFET-2015-21, EOS-2015-02}

\begin{abstract}
\vspace{0.2cm}\noindent
We set up a zero recoil sum rule to constrain the form factors of the
$\Lambda_b \to \Lambda_c$ transition. Our results are compared with the recent
lattice calculation for these transitions. We find the same situation as in the case for
$B \to D^*$: The lattice results practically saturate the sum rules, leaving basically no
room for excited states.
\end{abstract}

\maketitle

\section{Introduction}
\noindent
The precise determination of the CKM matrix elements $V_{xb}$, $x = c,u$
becomes increasingly important as an input for tests of the standard model at
the precision level. Although lattice QCD as well as non-lattice methods --
such as QCD sum rules -- have made enormous progress, we are still facing a
tension between determinations of $V_{xb}$ from inclusive versus exclusive
decays \cite{PDGReview}.

It is generally believed that $|V_{cb}|$ can currently be determined with the
best precision via the inclusive decay $B \to X_c \ell \bar{\nu}$
\cite{Benson:2003kp,Gambino:2013rza}. In this case
one applies an operator product expansion (OPE) in terms of local operators,
which sets up and expansion for the total rate, as well as for spectral
moments, in powers of $\alpha_s$ and $\Lambda_{\rm QCD} / m_Q$, $Q = b,c$. This
combined expansion seems to converge rapidly, giving us confidence in the precision of the method.

On the other hand, exclusive
decays also allow for a precise determination of $V_{cb}$ from the decays $B
\to D^{(*)} \ell \bar{\nu}$ by extrapolating to the point of maximal momentum
transfer to the leptons \cite{PDGReview}. At this point, heavy quark symmetries yield an
absolute normalization of the form factors, and corrections to the form factor
normalizations can be computed on the lattice \cite{Bailey:2014tva,Lattice:2015rga}
as well as from QCD sum rules \cite{Bigi:1994ga,Kapustin:1996dy,Gambino:2010bp,Gambino:2012rd}.

The aforementioned tension between the inclusive and the exclusive
determinations of $|V_{cb}|$ is driven by the lattice values for the form
factor normalization for the $B \to D^{(*)}$ form factors. On the other hand,
the anatomy of the $b \to c$ transition at zero recoil can be studied with
zero-recoil sum rules, which hint at
smaller values for the form factor normalizations and which are fully consistent
with the inclusive determination. In particular, from the point of sum rules,
the current lattice value would imply unexpectedly small contributions from the
excited states \cite{Gambino:2010bp,Gambino:2012rd}.

More serious seems the problem with the determinations of $|V_{ub}|$. The
inclusive determination relies on a light-cone version of the
OPE leading to the corresponding heavy mass expansion \cite{PDGReview}. The hadronic
input -- the so called shape functions -- are not well known (in particular at subleading
order), and thus the resulting expansion leads to larger uncertainties compared to ones in the local OPE
relevant for semileptonic $b\to c$ decays.

The exclusive determinations on $V_{ub}$
rely mainly on the channel $B \to \pi \ell \bar{\nu}$.  For this decay, the form
factors need to be computed either on the lattice \cite{Lattice:2015tia} or
estimated via light-cone sum rules \cite{Khodjamirian:2011ub}.
Using these form factors, which turn out to be consistent between the lattice
and the QCD sum rules, a value of
$|V_{ub}|$ can be extracted that is about three standard deviations smaller
than the inclusive one.

Since currently the exclusive determination of $V_{ub}$ rests mainly on a single
channel, it is important to have an independent determination from an other
channel. Since the purely leptonic decay $B \to \ell \bar{\nu}$ suffers - even
for the $\tau$ lepton - from helicity suppression, the existing measurements of
$B \to \tau \bar{\nu}$ are currently too imprecise to decide between the exclusive and
inclusive value of $V_{ub}$. This tension has also lead to sepeculations (see e.g. \cite{Crivelin:2014zpa,Buras:2010pz}
that ``new physics'' is responsible for the effect, although right-handed currents
have recently been excluded as an explanation \cite{Feldmann:2015xsa}.

Recently the LHCb collaboration published a first measurement of the branching
ratio of $\Lambda_b \to p \ell \bar{\nu}$ \cite{Aaij:2015bfa}, which is in
principle precise enough to challenge determinations based on $B\to
\pi\ell\bar\nu$. However, this measurement is normalized to the branching ratio
of $\Lambda_b \to \Lambda_c \ell \bar{\nu}$. Thus, the extraction of the ratio
$|V_{ub} / V_{cb}|$ requires the form factors to be calculated for both the
$\Lambda_b \to p$ as well as for the $\Lambda_b \to \Lambda_c$ transition. This
has been done recently on the lattice for both transitions with sufficient precision in
\cite{Detmold:2015aaa}. Their results for the $\Lambda_b\to p$ transitions
compare favorably with light-cone sum rule calculations \cite{Mannel:2011xg}, but the
precision of these sum rules is intrinsically limited.

In this work, we construct a zero-recoil sum rule (ZRSR) for the $\Lambda_b
\to \Lambda_c$ transitions, along the same lines as for the  $B \to D^*$ form
factor, see e.g. \cite{Gambino:2010bp}.  We shall investigate in this paper, if
the tension present in the lattice calculation versus the zero-recoil sum rule
for the mesons persists for the case of the baryons. In the next section we
formulate the zero-recoil sum rule for baryons and compute the necessary OPEs
to the required level of precision. We apply this method to both the axial
vector and the vector current, which eventually yields constraints for a subset of the
from factors that describe the $\Lambda_b \to \Lambda_c$ transitions.
Finally we compare our results with the lattice values and conclude.

\section{Zero Recoil Sum Rule}
\noindent
The sum rule at zero recoil ist set up in the same way as in the case for
mesons \cite{Gambino:2010bp} by considering the forward matrix element
\begin{equation}
    T_\Gamma(v \cdot q ) \equiv \frac{1}{N_\Gamma} \int d^4 x \, e^{ -i \, (v\cdot x)  \, (v \cdot q)}
    \langle  \Lambda_b (P) | \mathcal T \big\{ \bar{b}(x) \Gamma c(x),\, \bar{c}(0) \Gamma b(0)  \big\} | \Lambda_b (P) \rangle
\end{equation}
where we shall discuss two possible choices of the currents $\Gamma \otimes
\Gamma$: $\gamma_\mu \otimes \gamma^\mu$  $(V \times V)$ and
$\gamma_\mu \gamma_5 \otimes \gamma^\mu \gamma_5$  $(A \times A)$.  The
normalization constants $N_\Gamma$ are chosen to be $N_V = 1$ and $N_A = 3$
for the two cases respectively.
Furthermore, $P\equiv M_{\Lambda_b} v$ is the momentum of the $\Lambda_b$
baryon, from which we define the velocity $v$.

\begin{figure}[h]
    \centering
    \includegraphics[scale=0.8]{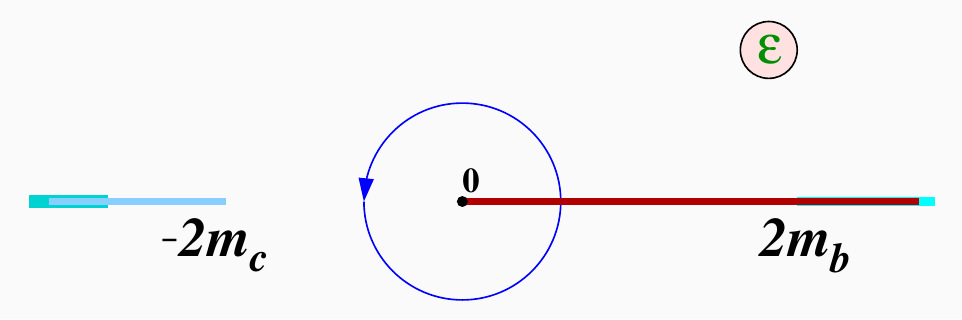}
    \caption{Integration contour for the calculation of $ I_{n,\Gamma}(\eps_M)$; the radius of the contour is $\eps_M$. Figure taken from \cite{Gambino:2010bp}.
    }
    \label{fig:contour}
\end{figure}

We want to set up a sum rule at the kinematical point where the charm quark also moves with the same velocity
$v$ which is the point of zero-recoil transferred by the $b \to c$ transition. Thus
we redefine the quark fields as
\begin{equation}
    \bar{b}(x) = e^{+i m_b (v \cdot x)} \bar{b}_v (x), \qquad\text{and}\qquad  c(x) = e^{-i m_c (v \cdot x)} c_v (x)\,,
\end{equation}
which suggests to define the parameter $\eps = m_b - m_c - (v\cdot q)$. We can
then reparametrize the forward matrix element in terms of $\eps$, which leads
to
\begin{equation}
    \label{eq:Teps}
    T_\Gamma (\eps) =  \frac{1}{N_\Gamma} \int d^4 x \, e^{ i \,  \,  (v\cdot x) \eps}
    \langle  \Lambda_b (P) | \mathcal T \big\{ \bar{b}_v(x) \Gamma c_v(x) \, \bar{c}_v(0) \Gamma b_v(0)  \big\} | \Lambda_b (P) \rangle
\end{equation}
Since $M_{\Lambda_b} - M_{\Lambda_c} \simeq m_b - m_c$, the quantity $\eps$
corresponds to the excitation energy of the intermediate charm states above the
$\Lambda_c$.
The steps leading to the sum rule are formally as in
\cite{Gambino:2010bp}, however, the relevant hadronic matrix elements will be different.
Along the lines of \cite{Gambino:2010bp} we define the contour integrals
\begin{equation}
    I_{n,\Gamma}(\eps_M) \equiv \frac{-1}{2\pi i} \oint_{|\eps| = \eps_M}  \, \eps ^n  \,
     T_\Gamma(\eps)\,\mathrm{d}\eps\, ,
\end{equation}
where the relevant contour is shown in \reffig{contour}.

Inserting a complete set of states, the lowest possible state is the $\Lambda_c$ moving with velocity $v$,
the higher states will excited states of the $\Lambda_c$ but also non-resonant contributions such as $\Lambda_c \pi$
or $D p$, where the charmed hadron moves with velocity $v$. Looking first at the
integral $I_{0,\Gamma}(\eps_M)$ the lowest contribution thus is related to the square
of the $\Lambda_b \to \Lambda_c$ matrix elements at zero recoil
\begin{equation}
    F
         \equiv \frac{1}{N_V}  \sum_{s'}
          \langle \Lambda_b (v,s) | \bar{b}_v \gamma_\mu  c_v | \Lambda_c (v,s') \rangle \, \langle \Lambda_c (v,s')  | \bar{c}_v \gamma^\mu  b_v | \Lambda_b (v,s) \rangle
\end{equation}
for the vector current, and
\begin{equation}
    G
         \equiv \frac{1}{N_A}  \sum_{s'}
          \langle \Lambda_b (v,s) | \bar{b}_v \gamma_\mu\gamma_5  c_v | \Lambda_c (v,s') \rangle \, \langle \Lambda_c (v,s')  | \bar{c}_v \gamma^\mu\gamma_5  b_v | \Lambda_b (v,s) \rangle\\
 \end{equation}
for the axial-vector current.

We use the form factors for the $\Lambda_b\to \Lambda_c$ transitions in the
helicity basis, which is introduced in \cite{Feldmann:2011xf}. For the vector
current they read
\begin{align}
\langle \Lambda_c (v',s')  | \bar{c} \gamma_\mu b | \Lambda_b (v,s)  \rangle
    & = \bar{u}_{\Lambda_c} (v',s')
        \left[ f_0 (w) ( M_{\Lambda_b} -   M_{\Lambda_c} ) \frac{q^\mu}{q^2}  \right.\\
    & + f_+ (w) \frac{M_{\Lambda_b} +  M_{\Lambda_c} }{s_+}
        \left( M_{\Lambda_b} v_\mu + M_{\Lambda_c} v_\mu^\prime  -  ( M_{\Lambda_b}^2 -   M_{\Lambda_c}^2 ) \frac{q^\mu}{q^2}  \right)\\
    &   \left.  + f_\perp (w) \left(\gamma_\mu - \frac{2 M_{\Lambda_c} M_{\Lambda_b}  }{s_+}  (v_\mu + v_\mu^\prime)  \right) \right]
        u_{\Lambda_b} (v,s)\,,
\end{align}
and for the axial vector current one has
\begin{align}
\langle \Lambda_c (v',s')  | \bar{c} \gamma_5 \gamma_\mu  b | \Lambda_b (v,s)  \rangle
    & = -\bar{u}_{\Lambda_c} (v',s') \gamma_5
        \left[ g_0 (w) ( M_{\Lambda_b} +   M_{\Lambda_c} ) \frac{q^\mu}{q^2}  \right.\\
    & + g_+ (w) \frac{M_{\Lambda_b} -   M_{\Lambda_c} }{s_-}
        \left( M_{\Lambda_b} v_\mu + M_{\Lambda_c} v_\mu^\prime  -  ( M_{\Lambda_b}^2 -   M_{\Lambda_c}^2 ) \frac{q^\mu}{q^2}  \right)\\
    &   \left.  + g_\perp (w) \left(\gamma_\mu + \frac{2 M_{\Lambda_c} M_{\Lambda_b}  }{s_+}  (v_\mu - v_\mu^\prime)  \right) \right] u_{\Lambda_b} (v,s)\,.
\end{align}
In terms of the heavy hadron velocities $v$, $v'$ and their scalar product $w = vv'$ one finds
$q =  M_{\Lambda_b} v - M_{\Lambda_c} v'$ and
$q^2 = M_{\Lambda_b}^2 +  M_{\Lambda_c}^2 - 2 M_{\Lambda_b}  M_{\Lambda_c} w$.
In addition, we abbreviate
\begin{equation}
s_\pm = (M_{\Lambda_b} \pm   M_{\Lambda_c})^2 - q^2\,.
\end{equation}

With these definitions we obtain
\begin{eqnarray}
    F &=& |f_0(w=1)|^2\,,\quad\text{and} \\
    G &=&  \frac{1}{3}\left[2|g_\perp(w=1)|^2 + |g_+(w=1)|^2\right]\,.
\end{eqnarray}

The form factors $f_\lambda$ and $g_\lambda$, $\lambda = 0,+,\perp$, have been
recently calculated on the lattice \cite{Detmold:2015aaa}, and are published in
form of a handful of parameters, including their correlation matrix. Using
their results for the form factors, the authors of \cite{Detmold:2015aaa}\footnote{
The values shown here are taken from the arXiv version 3.
}
obtain at the zero recoil point $w = 1$:
\begin{equation}
    \label{eq:lattice-results}
    F  = 0.972 \pm 0.058\,,\qquad\text{and}\qquad G  = 0.817 \pm 0.044\,.
\end{equation}

In the rest of this paper we confront the above lattice results with the
constraints obtained form the zero-recoil sum rule

\subsection{Axial Vector Sum Rule at Zero Recoil}
\label{sec:zrsrA}

\begin{table}
\centering
\renewcommand{\arraystretch}{1.2}
\begin{tabular}{|c| c| c| c| c|}
    \hline
    Parameter                                   & mean value/ $1\sigma$ interval & unit     & prior               & source/comments\\
    \hline
    \multicolumn{5}{|c|}{quark-gluon coupling and quark masses}\\
    \hline
    $\alpha_s(m_Z)$                             &  0.1184  $\pm$ 0.0007          & ---      & gaussian $@$ $68\%$ &  \cite{Agashe:2014kda}  \\
    $\mbMSbar$                                  &    4.18  $\pm$ 0.03            & \GeV     & gaussian $@$ $68\%$ &  \cite{Agashe:2014kda}  \\
    $\mcMSbar$                                  &    1.275 $\pm$ 0.025           & \GeV     & gaussian $@$ $68\%$ &  \cite{Agashe:2014kda}  \\
    \hline
    \multicolumn{5}{|c|}{hadronic matrix elements (nominal choice)}\\
    \hline
    $\mu^2_\pi(1\,\GeV)$                        &    0.50  $\pm$ 0.10            & $\GeV^2$ & gaussian $@$ $68\%$ &  see \refeq{mu2pi} \\
    $\rho^3_D(1\,\GeV)$                         &    0.17  $\pm$ 0.08            & $\GeV^3$ & gaussian $@$ $68\%$ &  see \refeq{rho3D} \\
    \hline
\end{tabular}
\caption{Summary of the prior PDFs used in the numeric analyses. The central values lead to
$m_b^\text{kin}(\mu = 0.75\,\GeV) = 4.62\,\GeV$, $m_c^\text{kin}(\mu = 0.75\,\GeV) = 1.20\,\GeV$,
and $\alpha_s(\sqrt{m_b^\text{kin} m_c^\text{kin}}) = 0.284$.}
\label{tab:priors}
\end{table}

\begin{figure}
    \centering
    \includegraphics[width=.49\textwidth]{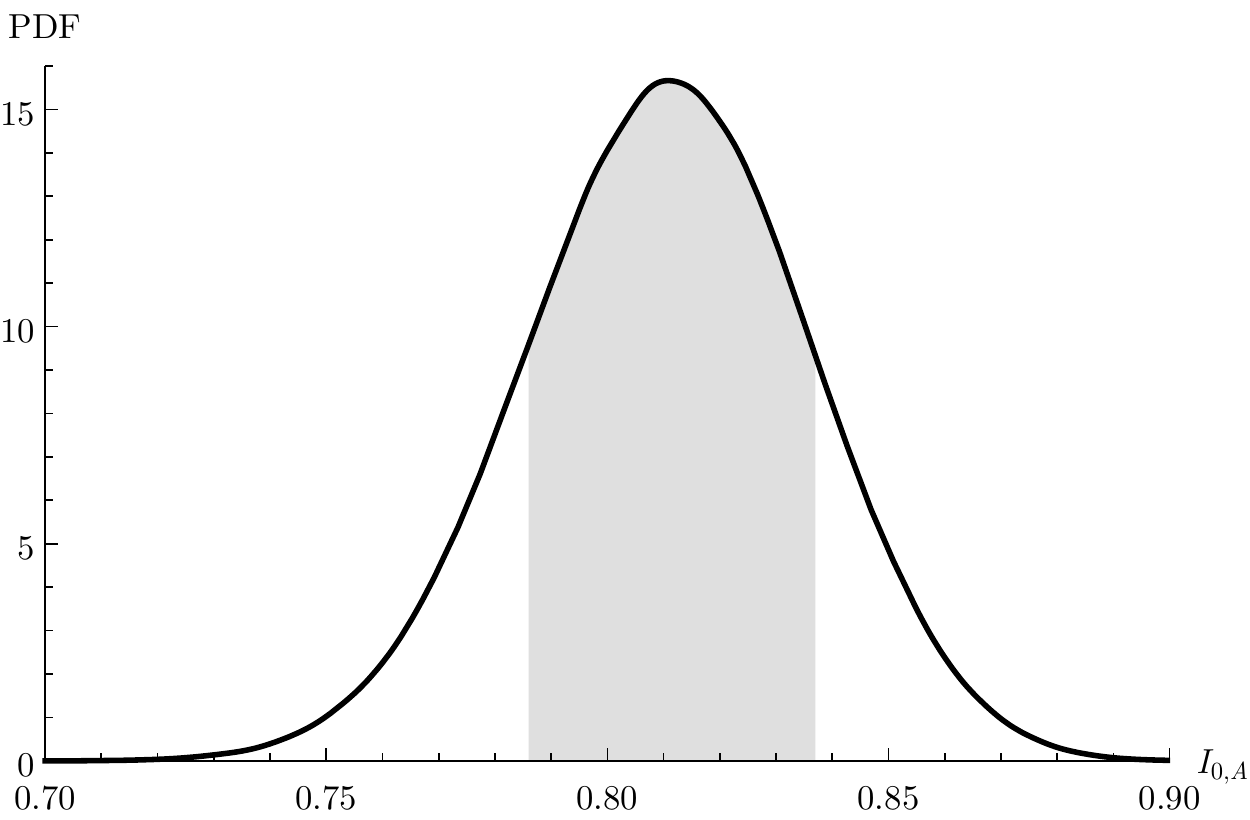}
    \includegraphics[width=.49\textwidth]{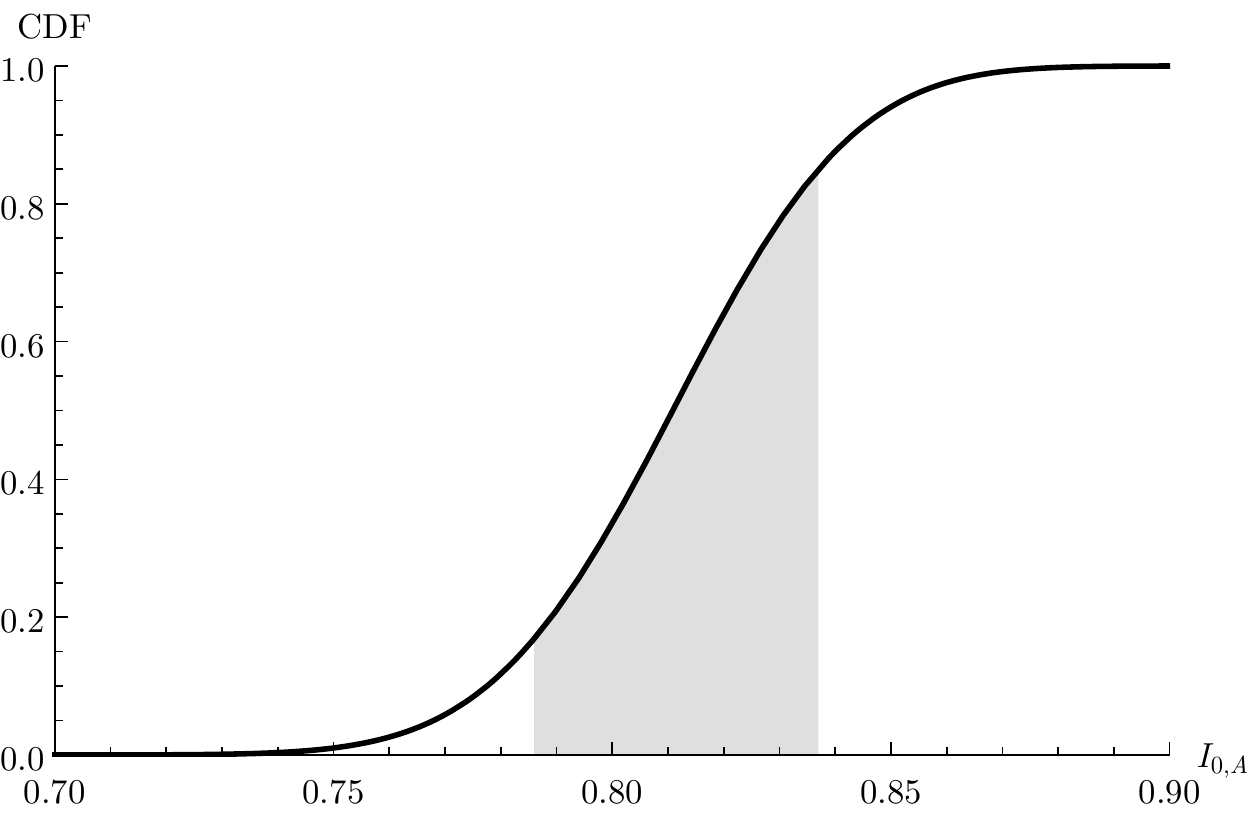}
    \caption{The result PDF (left) and the CDF (right) for the quantity $I_{0,A}$ as
        obtained from $10^6$ random samples of the paameter space. We show the
        central $68\%$ probability interval as the grey-shaded area.
    }
    \label{fig:I0A}
\end{figure}

\noindent
We start the discussion with the axial vector sum rule
\begin{equation}
\label{eq:zrsrA}
\begin{aligned}
    I_{0,A}(\eps_M)
        & = \frac{1}{N_A} \sum_{X_c,\,\eps \le \eps_M}
          \bra{\Lambda_b(v,s)} \bar{b}_v \gamma_\mu \gamma_5 c_v\ket{X_c(v)}
          \bra{X_c(v)} \bar{c}_v\gamma^\mu \gamma_5 b_v\ket{\Lambda_b(v,s)}\\
        & \equiv G  + G_\text{inel}(\eps_M)
\end{aligned}
\end{equation}
In the above, $G_\text{inel}(\eps_M)$ captures all inelastic contributions to
the correlation function up to an energy $\eps_M$, i.e., all contributions with excitation energies
$0 < \epsilon \le \epsilon_M$.  Note that both terms $G$ and
$G_\text{inel}(\eps_M)$ are positive. We can therefore rewrite the sum rule as an
upper bound for   $G$:
\begin{equation}
    G \leq I_{0,A}(\eps_M)\,.
\end{equation}

The left-hand side of \refeq{zrsrA} can be evaluated in the OPE
\cite{Gambino:2010bp}, and one obtains
\begin{equation}
    I_{0,A} (\epsilon_M) = \xi^{\rm pert}_A(\epsilon_M,\mu)
    - \Delta^A_{1/m^2}(\epsilon_M,\mu) - \Delta^A_{1/m^3}(\epsilon_M,\mu)
    + \order{\Lambda_\text{had}^4/m_b^4, \Lambda_\text{had}/m_c^4}
\end{equation}
where the perturbative contribution is the same as for the mesonic case
\begin{equation}
    \xi^{\rm pert}_A(\eps_M = \mu = 0.75\,\GeV) = 0.970  \pm 0.02 \,,
\end{equation}
which contains the $\alpha_s$ and the $\alpha_s^2$ corrections \cite{Gambino:2010bp}.

The power corrections differ from the mesonic results, since a priori the
forward matrix elements for the $\Lambda_b$ are different from the ones for the
$B$ mesons. Furthermore, for the $\Lambda$-like heavy baryons, the matrix
elements of all the spin-triplet operators vanish.  This is due to the fact
that the light degrees of freedom do not have any angular momentum and thus
cannot generate a chromomagnetic field. Hence, all matrix elements involving
these operators -- including $\mu^2_G(\Lambda_b)$, $\rho^3_{LS}(\Lambda_b)$ --
vanish. The non-perturbative power corrections for the baryonic case therefore
read
\begin{align}
    \Delta^A_{1/m^2} & = \frac{\mu_\pi^2 (\Lambda_b) }{4} \left( \frac{1}{m_c^2} +  \frac{1}{m_b^2} + \frac{2}{3 m_b m_c }\right)  \\
    \Delta^A_{1/m^3} & = \frac{\rho_D^3 (\Lambda_b) }{4 m_c^3} +
                                \frac{\rho_D^3 (\Lambda_b) }{12 m_b}   \left( \frac{1}{m_c^2} +  \frac{3}{m_b^2} + \frac{1}{ m_b m_c }\right).
\end{align}

The kinetic energy operator for the $\Lambda_b$ baryon has been discussed in
the context of the $\Lambda_b$ baryon lifetime \cite{Neubert:1996we}. Using the
spin-averaged heavy meson masses one obtains up to terms of order $1/m$
\begin{equation}
\label{eq:delta-mu2pi}
    \mu_\pi^2(B) - \mu_\pi^2(\Lambda_b) = \frac{2 m_b m_c}{m_b - m_c} \left((M_{\Lambda_b} - M_{\Lambda_c}) - (\overline{M}_B - \overline{M}_D)\right)
    (1+ \mathcal{O}(1/m^2))\,.
\end{equation}
The most recent results a of combined fit of the $B$-meson hadronic matrix
elements and $V_{cb}$ to the measured lepton-energy moments in
$B\to X_c \ell\nu$ yield $\mu_\pi^2(B) = (0.47 \pm 0.07)\,\GeV^2$
\cite{Alberti:2014yda}.  Using \refeq{mu2pi} this translates to
\begin{equation}
\label{eq:mu2pi}
    \mu_\pi^2(\Lambda_b) = (0.50 \pm 0.10)\,\GeV^2\,,
\end{equation}
where we increase the uncertainty to account for the lack of $1/m^2$ terms.

Given the small difference between the kinetic energy parameters of baryons and mesons, we use also for the
Darwin term of the $\Lambda_b$  the same value as for the $B$-meson.  The mesonic matrix element
is obtained in \cite{Alberti:2014yda}; for the  $\Lambda_b$ we use the same central value and
increase the uncertainty by a factor of two,
\begin{equation}
\label{eq:rho3D}
    \rho_D^3(\Lambda_b) \simeq (0.17 \pm 0.08)\,\GeV^3\,.
\end{equation}

Using these numbers,  we obtain
\begin{align}
    \Delta^A_{1/m^2}  (\eps_M = \mu = 0.75\,\GeV) & = 0.108 \,,\\
    \Delta^A_{1/m^3}  (\eps_M = \mu = 0.75\,\GeV) & = 0.028 \,.
\end{align}
We note that $\Delta_{1/m^2}$ is  about $  20\%$ larger than for the mesonic case,
while $\Delta_{1/m^3}$  for the $\Lambda_b$ baryon yields numerically the same result as for
the meson. The above results shall only be illustrative, and have been obtained for our default
choice of input parameter.

For a more thorough numerical study, we use and extend EOS \cite{EOS}. This allows us
to carry out a Bayesian uncertainty propagation based on Monte Carlo techniques.
We choose prior probability density functions
(PDFs) for all input parameters based on the principle of maximum entropy
\cite{Jaynes:1957zza}. We use Gaussian distributions throughout this work, since
in all cases the mean and variance of the parameters are known.
For a summary of the PDFs, see \reftab{priors}. We draw $10^6$ random samples
from $P(I_{0,A})$, the PDF of our quantitiy of interest. The result PDF and the
corresponding Cumulative Probability Density Function (CDF) are shown in
\reffig{I0A}. For our choice of the prior PDFs, the result is a gaussian
distribution to very good accuracy, with skewness $-0.08$ and excess kurtosis
of $-0.04$. From the result PDF we obtain the mode and the central $68\%$
probability interval
\begin{equation}
    I_{0,A}(\eps_M = \mu = 0.75) = 0.811^{+0.025}_{-0.026} \,.
\end{equation}

\begin{figure}
    \centering
    \includegraphics[width=.49\textwidth]{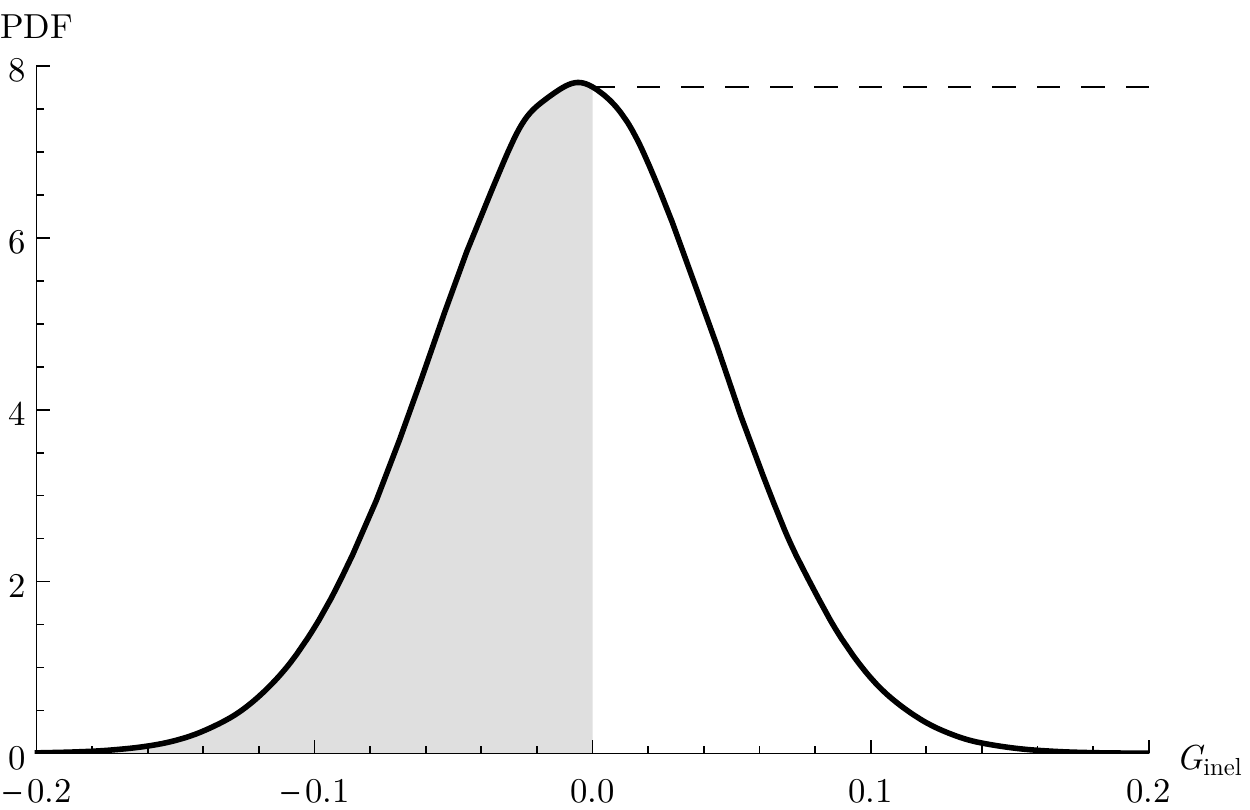}
    \includegraphics[width=.49\textwidth]{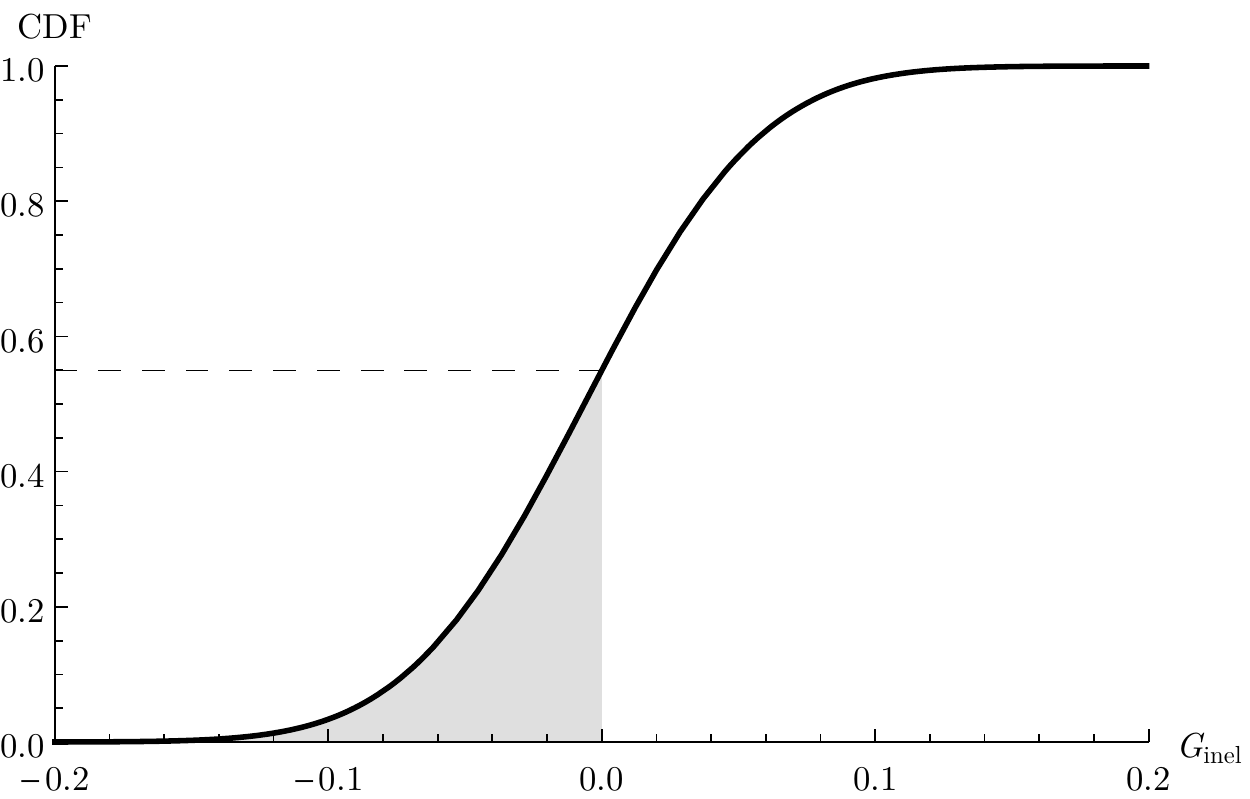}
    \caption{The result PDF (left) and CDF (right) of $G_\text{inel}$ as obtained from $10^6$ samples
        of both the OPE result for $I_{0,A}$ and the lattice result for $G$.
        The grey-shaded area shows those results with $G_\text{inel} < 0$, which
        is unphysical. We find that the unphysical range of $G_\text{inel}$
        accumulates $\sim 55\%$ probability.
    }
    \label{fig:Ginel}
\end{figure}

Following \refeq{zrsrA}, we can also compute the inelastic contributions
$G_\text{inel}$ from our nominal results for $I_{0,A}$ and the lattice
results for $G$. Using $10^6$ samples of both quantities, we obtain the PDF
and CDF for the quantity $G_\text{inel}$ as shown in \reffig{Ginel}. Again,
the PDF is approximately gaussian with skewness of about $ -0.01$ and excess
kurtosis of about $  -0.006$. We obtain the mode of the distribution and the
central $68\%$ probability interval as
\begin{equation}
    \label{eq:GinelMean}
    G_\text{inel} = -0.005^{+0.049}_{-0.052}\,.
\end{equation}
Roughly $55\%$ of the samples of $G_\text{inel}$ turn out to be unphysical,
since they are negative. Thus we conclude from this statistical analysis that
the situation for the $\Lambda_b \to \Lambda_c$ is very similar as for the $B
\to D^*$ case: The lattice results for the form factors at zero recoil saturate
the corresponding zero-recoil sum rule by a very large degree, leaving almost
no room for inelastic contributions. In fact, compared to the mesonic case, the
situation seems to be even worse, since the central value obtained from the
lattice \refeq{lattice-results} exceeds the central value for our upper bound.
Furthermore, for the mesonic case, one may estimate the inelastic
contributions, which turn out to be sizable. This in turn implies that the
zero-recoil sum rules  would predict a smaller value for the form factors.
Unfortunately, the estimates in the mesonic case rely on the so-called BPS
limit, which cannot be used in the case of baryons. Since an estimate of the
inelastic contributions in the case of the $\Lambda_b$ requires (possible even
model dependent) input, we will not discuss this in the present paper.

\subsection{Vector Sum Rule at Zero Recoil}
\label{sec:zrsrV}
The vector sum rule is obtained from \refeq{Teps} by inserting
$\Gamma \otimes \Gamma = \gamma_\mu \otimes \gamma^\mu$ and $N_V = 1$,
\begin{equation}
\label{eq:zrsrV}
\begin{aligned}
    I_{0,V}(\eps_M)
        & = \frac{1}{N_V} \sum_{ X_c, \, \eps \le \eps_M}
          \bra{\Lambda_b(v,s)} \bar{b}_v \gamma_\mu c_v\ket{X_c(v)}
          \bra{X_c(v)} \bar{c}_v \gamma^\mu b_v\ket{\Lambda_b(v,s)}\\
        & \equiv F  + F_\text{inel}(\eps_M)\,.
\end{aligned}
\end{equation}
Analogous to the axial vector current, $F_\text{inel}(\eps_M)$ captures all
inelastic contributions to the correlation function with excitation energies less than $\eps_M$,
i.e.,  all contributions
with excitation energies $0 < \epsilon \le \epsilon_M$. Again, $F$ and
$F_\text{inel}(\eps_M)$ are positive, and we can therefore rewrite the sum rule as
an upper bound for the term $F$:
\begin{equation}
    F \leq I_{0,V}(\eps_M)\,.
\end{equation}

\begin{figure}
    \centering
    \includegraphics[width=.49\textwidth]{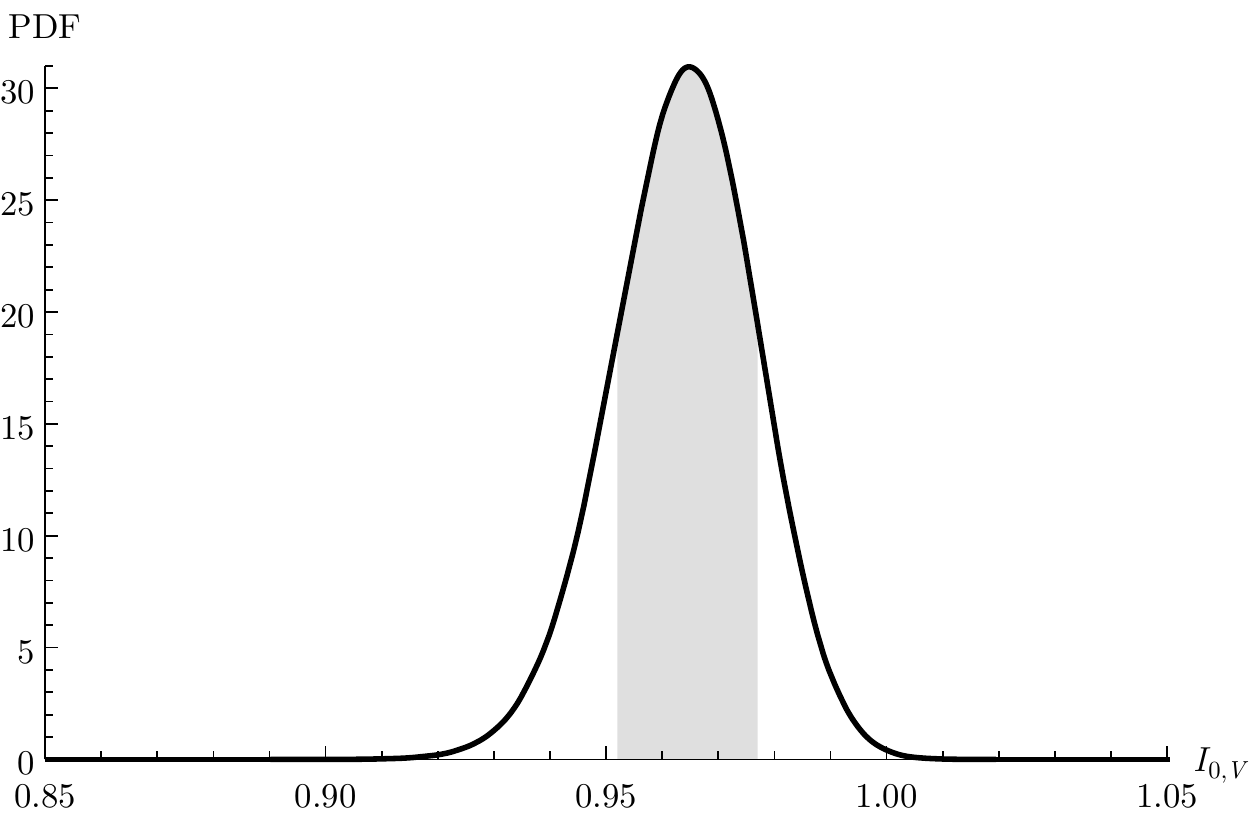}
    \includegraphics[width=.49\textwidth]{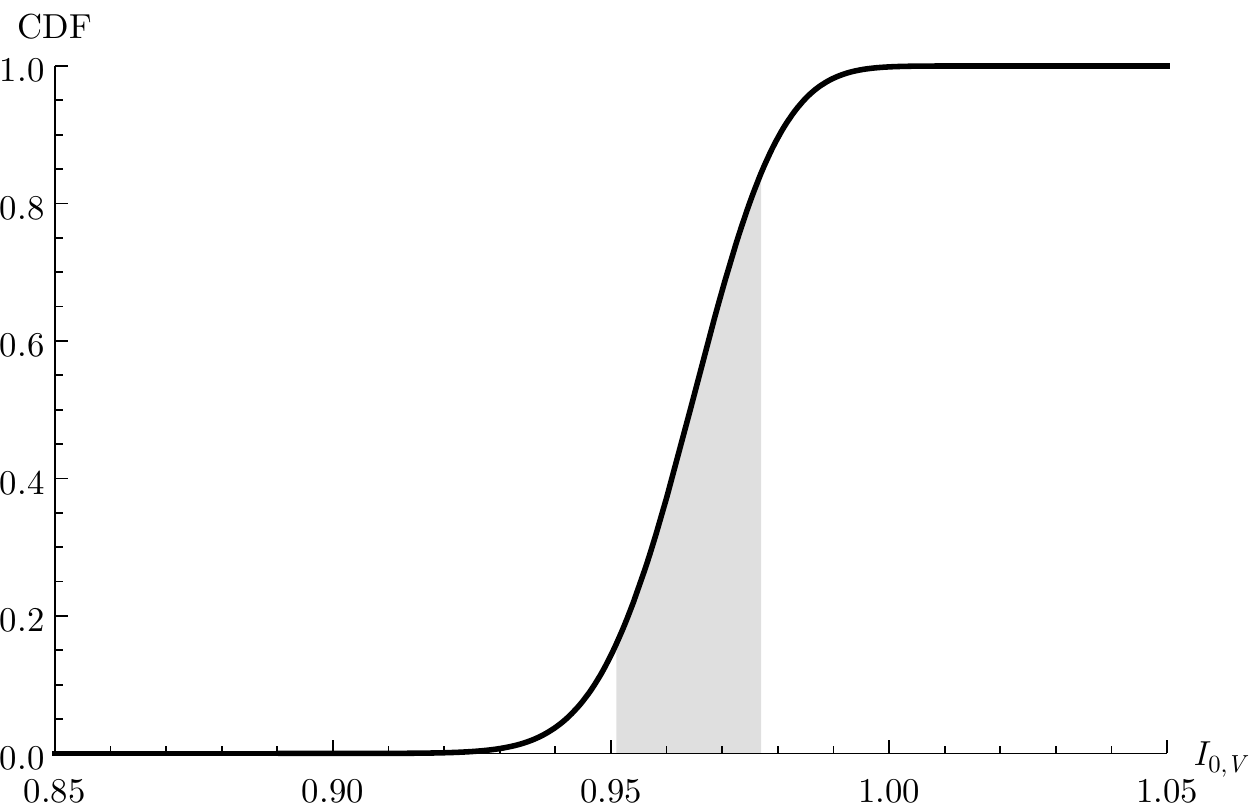}
    \caption{The result PDF (left) and CDF (right) for the quantity $I_{0,V}$ as
        obtained from $10^6$ random samples of the paameter space. We show the
        central $68\%$ probability interval as the grey-shaded area.
    }
    \label{fig:I0V}
\end{figure}

The OPE result for the left-hand side of \refeq{zrsrV} reads
\begin{equation}
    I_{0,V} (\epsilon_M) = \xi^{\rm pert}_V(\epsilon_M,\mu)
    - \Delta^V_{1/m^2}(\epsilon_M,\mu) - \Delta^V_{1/m^3}(\epsilon_M,\mu)
    + \order{\Lambda_\text{had}^4/m_b^4, \Lambda_\text{had}/m_c^4}
\end{equation}
where the perturbative contribution has been evaluated to order $\alpha_s$ in \cite{Uraltsev:2003ye}.
For the central values of the input parameters we obtain
\begin{equation}
    \xi^{\rm pert}_V(\eps_M = \mu = 0.75\,\GeV) = 1.03^{+0.03}_{-0.01} \,,
\end{equation}
where the uncertainty is estimated form a variation of the scale $0 \le \mu \le 1.5$ GeV.

The nonperturbative corrections have been given in \cite{Bigi:1994ga,Uraltsev:2003ye}
\begin{align}
    \Delta^V_{1/m^2} & = \frac{\mu_\pi^2 (\Lambda_b)}{4} \left(\frac{1}{m_c} - \frac{1}{m_b}\right)^2\\
    \Delta^V_{1/m^3} & = \frac{\rho_D^3 (\Lambda_b)}{4} \left(\frac{1}{m_c} - \frac{1}{m_b}\right)^2 \left(\frac{1}{m_c} + \frac{1}{m_b}\right)
\end{align}
and reflect the fact that the vector current is conserved in the limit $m_b = m_c$.

\begin{figure}
    \centering
    \includegraphics[width=.49\textwidth]{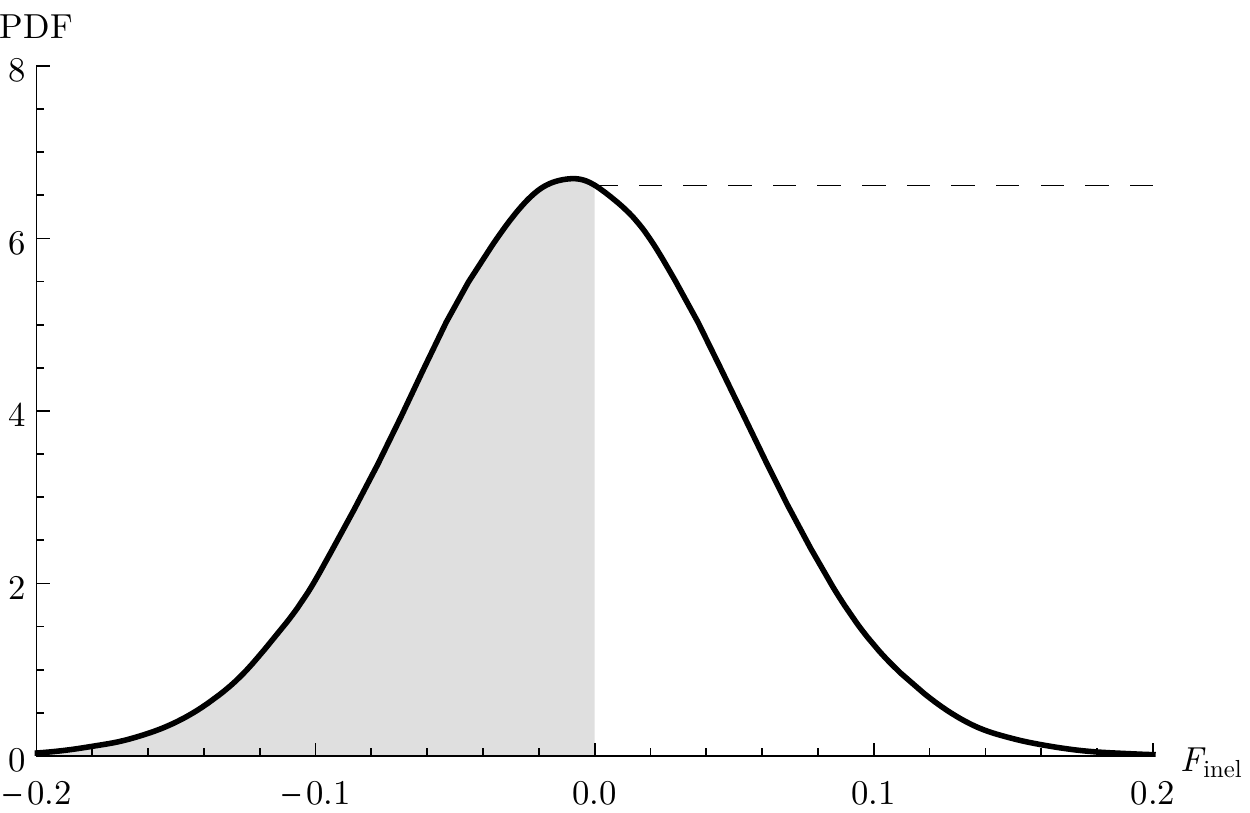}
    \includegraphics[width=.49\textwidth]{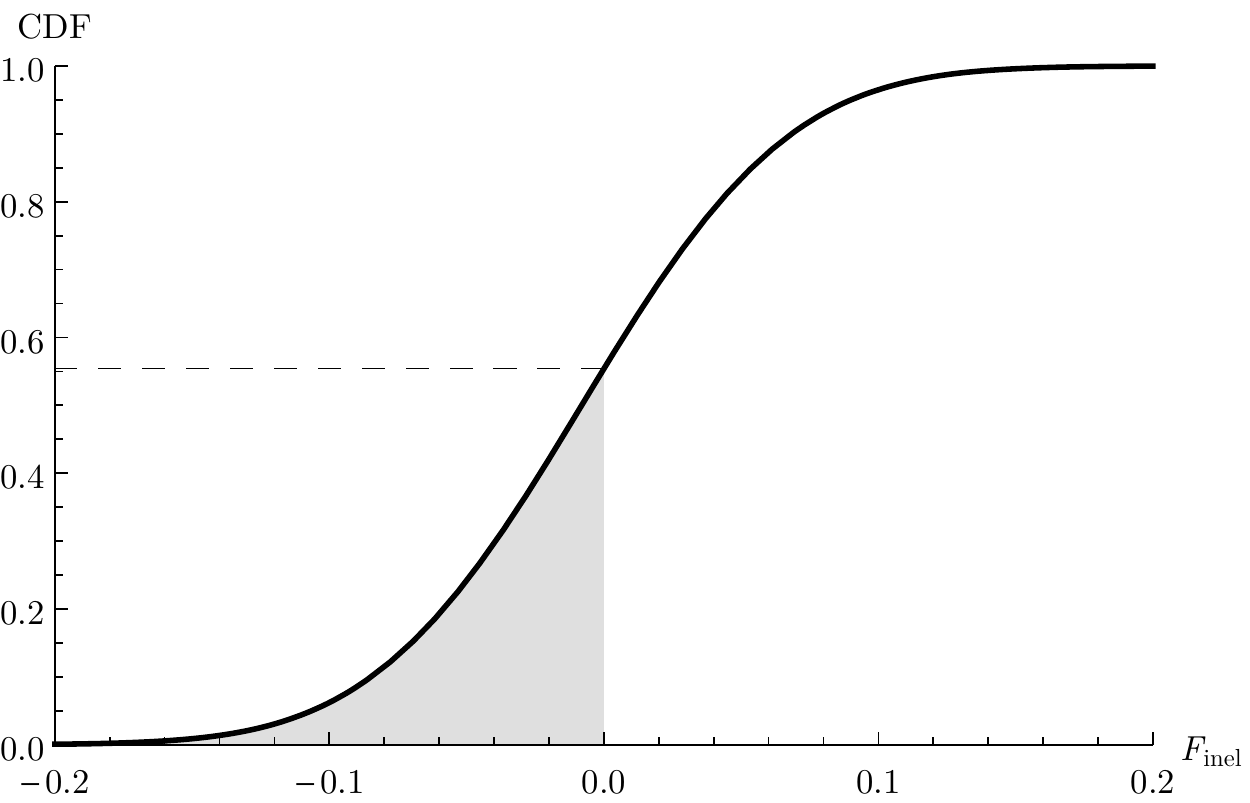}
    \caption{The PDF and the CDF of $F_\text{inel}$ as obtained from $10^6$ samples
        of both the OPE result for $I_{0,V}$ and the lattice result for $F$.
        The grey-shaded area shows those results with $F_\text{inel} < 0$, which
        is unphysical. We find that the unphysical range of $F_\text{inel}$
        accumulates $\sim 55\%$ probability.
    }
    \label{fig:Finel}
\end{figure}

Inserting the central values for the hadronic matrix elements, we obtain
\begin{align}
    \Delta^V_{1/m^2}  (\eps_M = \mu = 0.75\,\GeV) & = 0.047 \,,\\
    \Delta^V_{1/m^3}  (\eps_M = \mu = 0.75\,\GeV) & = 0.017 \,.
\end{align}
As before, these results are only meant as an illustration, and we repeat the
statistical procedure as outlined in \refsec{zrsrA}. We obtain for the mode and
central $68\%$ probability interval of the result PDF for $I_{0,V}$
\begin{equation}
    I_{0,V}(\eps_M = \mu = 0.75) = 0.965 \pm 0.013 \,,
\end{equation}
based on $10^6$ samples.  We display the resulting PDF and CDF for $I_{0,V}$ in
\reffig{I0V}. We compute the inelastic contribution as well -- just as before
in the case of the axialvector current -- and obtain
\begin{equation}
    \label{eq:FinelMean}
    F_\text{inel} = -0.010^{+0.061}_{-0.057}\,.
\end{equation}
as the mode and central uncertainty interval at $68\%$ probability; see
\reffig{Finel} for the respective result PDF and CDF. We further find that
$\sim 55\%$ of the drawn samples are unphysical, i.e., they show a negative
inelastic contribution.

Thus our findings are qualitatively the same as in the case of the axial current: The lattice result for the
scalar vector form factor $f_0$  at the non-recoil point again saturates the the zero-recoil sum rule to a
very large degree, leaving also for this case almost no room for inelastic contributions.

\section{Discussion and Conclusion}
\noindent
The determination of CKM matrix elements from exclusive semileptonic decays requires reliable calculations
for the form factors describing the corresponding hadronic transition. Since the form factors are genuinely
non-perturbative, the only known ``ab initio'' calculational method is lattice QCD. The progress in this field made in the last
years in the construction of efficient algorithms as well as the increasing computing power has turned lattice
calculation of form factors into an indispensable tool in flavor physics.

However, despite this progress it is important to perform checks of the lattice results from ``continuum'' methods.
One of these methods are QCD sum rules. On the one hand they are firmly rooted in QCD, on the other hand they allow
for a detailed study of the ``anatomy'' of the results obtained e.g.\ for form factors. It has to be clear that a QCD sum rule
can never make a precision prediction for a hadronic quantity, since the method is intrinsically limited to a level of a few
ten percent.

Nevertheless, QCD sum rules can serve to validate results obtained from other methods, e.g from lattice QCD.
In particular, the zero-recoil sum rules can give a hint on the sizes of the from factors at the non-recoil point; in case of the
$B \to D^*$ transition one can combine the  zero-recoil sum rule with an estimate for the inelastic contributions to actually
estimate the form factor itself.

In the analysis presented here we have shown that the lattice results \cite{Detmold:2015aaa}
for the $\Lambda_b \to \Lambda_c$ transition form factors saturate the zero-recoil sum rule to a large extent.
In fact, we found that the central values for the lattice results exceed the sum rule's upper bounds,
leaving practically no room for any inelastic contribution. This seems to be the case for both the axial-vector
as well as for the vector current.

In fact, the degree of saturation of the sum rule for the  $\Lambda_b \to \Lambda_c$ seems to be higher than for the
$B \to D^*$ transition, where the lattice value for the form factor at zero recoil still leaves room for a (too?) small
inelastic contribution.  Unfortunately, the inelastic contributions for the baryonic case are harder to estimate than in the
mesonic case; any estimate of the inelastic contributions for the baryons would require (probably model-dependent)
additional input. We leave the discussion of this to future work.

\acknowledgments
This work was supported by the German Minister for Education and Research (BMBF), contract No. 05H15PSCLA and
by the German Science Foundation (DFG) through the  DFG Research Unit FOR 1873 (``Quark
Flavour Physics and Effective Field Theories'').
We thank Stefan Meinel for helpful and rapid communication prior to submission of this
manuscript.

\bibliographystyle{apsrev4-1}
\bibliography{references.bib}
\end{document}